## Title

ARISA data from the human gut microbiome can detect individual differences observed by 454 sequencing regardless of binning strategy.

## Authors and Affiliations

Robert W. Reid[1], Melanie D. Spencer[2], Timothy J. Hamp[3], Anthony A. Fodor[3]

1- Bioinformatics Services Division, The University of North Carolina at Charlotte, 9201 University City Boulevard, Charlotte, NC 28223
2- Nutrition Research Institute, University of North Carolina, Kannapolis, North Carolina
3- Department of Bioinformatics and Genomics, The University of North Carolina at Charlotte, 9201 University City Boulevard, Charlotte, NC 28223

## Abstract

ARISA (Automated Ribosomal Intergenic Spacer Analysis) is a low-cost technique that allows for the rapid comparison of different microbial environments. In this study, we asked if a set of ARISA profiles can distinguish human microbial environments from one another with the same accuracy as results generated from 454 high throughput DNA sequencing. Using a set of human microbial communities where the sequencing results cluster by subject, we tested how choices made during ARISA data processing influence clustering. We found that choice of clustering methods had a profound effect with Ward's clustering generating profiles the most similar to 454 sequencing. Factors such as bin size, using presence or absence calls and technical replicate manipulation had a negligible effect on clustering. In fact, no established bin sizing method reported in the literature performed significantly different results than simply picking bin intervals at random. We conclude that in an analysis of ARISA data from an ecosystem of sufficient complexity to saturate bins, a careful choice of clustering algorithm is essential whereas differing strategies for choosing bins are likely to have a much less pronounced effect on the outcome of the analysis. As a tool for distinguishing complex microbial communities, ARISA closely approximates the results obtained from DNA sequencing at a fraction of the cost; however ARISA fails to reproduce the sequencing results perfectly.

## Introduction

Since the vast majority of microbial species within a given environment are not amenable to cell culture [1,2,3], molecular biology techniques have been developed to identify taxa based on genetic makeup [4,5]. While recent breakthroughs in "next generation" DNA sequencing technology have substantially decreased costs, sequencing can still be expensive, often limiting the breadth and depth of sample size of metagenomic surveys. Alternatives to direct DNA sequencing include DNA fingerprinting techniques such as terminal restriction fragment length polymorphisms (T-RFLP)[6], ARDRA[7], DGGE[8], and 2D-PAGE[9].

In this study, we focus on the automated method of ribosomal intergenic spacer analysis (ARISA) [10,11], a molecular biology technique derived from RISA, first described by Borneman and Triplett [12]. ARISA determines the structure of the microbial community by PCR amplifying the intergenic region between the 16S and 23S rRNA genes. ARISA can provide a rapid profile of an entire microbial community at a very low cost compared to DNA sequencing. In the generation of an ARISA profile, DNA from a community is isolated and the intergenic regions are PCR amplified. The resulting DNA fragments are separated via capillary electrophoresis according to size and each fragment length can be estimated from known size standards that are concurrently run along with the DNA fragments.

When distinguishing different microbial communities via ARISA, there are many choices during data processing and clustering that can potentially influence the results. We compared different parameters involved in ARISA data processing in an effort to best differentiate one microbial environment from another. A common analytical strategy is to group neighboring data signal into bins and assign an appropriate nucleotide length based on size standards. The sizes of these bins vary and there have been numerous binning strategies employed in the literature (Table 1). To date there has been no systematic exploration comparing these different binning strategies.

**Table 1: Summary of recent articles and the variety of bin sizes used in analysis.**

| Article | Bin Size (nt = nucleotide) |
|---|---|
| Soo et al., 2009[38] | Simple bin of 2 nt |
| Popa et al., 2009[13] | Calculated fragment length based on average and variability of technical replicates |
| Li et al., 2008[14] | Calculated fragment length based on average of 3 technical replicates |
| Ramette, 2009[39] | Shifting bin method [23] |
| Denman et al., 2008[40] | Simple bin of 2 nt |
| Wood et al., 2008[41] | Simple bin of 2 nt |
| Wood et al., 2008[42] | Simple bin of 3 nt |
| Lear et al., 2008[43] | Simple bin of 1 nt |

To test the various parameters, we used ARISA profiles from a human subject time course study[15], for which the microbial community composition has been confirmed independently by 454 sequencing. This study sampled repeatedly over time 15 subjects fed a rigorously controlled experimental diet. Our 454 sequencing demonstrated that the microbial communities of the human gut group clearly by subject over the 60 day time-course of our study despite subjects being fed the identical experimental diet[15]. Given this baseline result that subjects have a distinct gut microbial community over time, we tested various ARISA parameters that yielded the best congruence between the ARISA and DNA sequencing results. Of the parameters influencing the clustering of ARISA profiles, it was the clustering method itself that most affected the outcome. Our results demonstrate that for a fraction of the cost, ARISA profiling can produce very similar results to 454 sequencing of 16S rRNA amplicon tags in profiling a human derived microbial community.

**Results**

A common use of ARISA is to cluster the ARISA fingerprints to determine similarities between different microbial communities. Figure 1 summarizes choices that can be made during the workflow for a set of ARISA profiles highlighting options (ovals) that can be made during analysis. We evaluated each of the options within an oval to determine how these choices affect the performance of clustering algorithms.

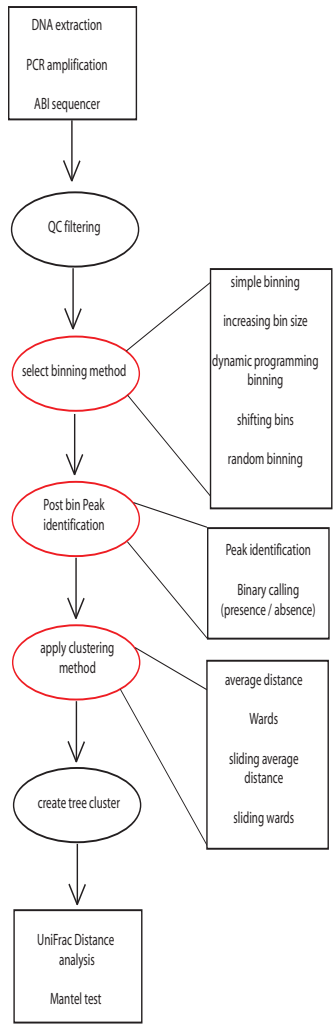

**Figure 1: Workflow for ARISA analysis used in the current study.** DNA is first extracted from the sample in question, PCR amplified, and then fragments are separated on a genetic analyzer. QC filtering techniques can be applied to identify poor runs. Data signals are converted into nucleotide length, and then converted into fractions of total intensity or binary format. Technical replicates are handled prior to binning peaks via three different strategies. Binned datasets are compared via a clustering method and dendrograms are created. Each cluster is compared to the model cluster based on 16S ribosomal gene region DNA sequencing using UniFrac. Each of the steps (ovals) has multiple options, which in this paper were tested for clustering performance.

## DNA sequencing

An ideal evaluation of algorithms that cluster ARISA data would utilize a dataset in which the expected outcome is known. In this paper, we take advantage of a large dataset of human gut microbiome samples for which we have both the ARISA results and the 16s rRNA sequences generated from 454 sequencing. This dataset was generated as part of a choline depletion study where subjects were placed on a tightly controlled diet over a 60 day time course to study the effects of choline depletion on the body[15]. All subjects within the study were placed on identical diets, stool samples were periodically collected, DNA was then extracted, and 16S rRNA gene sequencing was performed to determine how gut microbial communities are influenced by diet. Multiple time points were taken over the course of the study, before choline depletion, during and after repletion. Both ARISA and DNA sequencing results were obtained for each time point for each subject in the study.

Sequencing for all time points for each subject was undertaken using 454 sequencing technology. Primers were selected to target the V1-V2 region of the 16S rRNA gene, ~200,000 DNA sequences were collected and assigned to an OTU with 97% similarity. The top 200 most commonly occurring OTUs were selected across the entire sequencing dataset for comparing time points and subjects. For each individual time point, the number of sequence reads for each of the 200 OTUs was tabulated. All time points across all subjects were then correlated with one another and clustered via Wards clustering method in order to classify profiles and determine which time points have similar OTU profiles. Figure 2A depicts the results of the hierarchical clustering procedure for all of the time points within the choline depletion study. As we might expect from another study on the gut human microbiome where over time diet was held constant (Fig. 3A in [18]) nearly all of the samples cluster by subject across time points. Across multiple cohorts, therefore, gut microbial differences between individuals appear to persist over time despite subjects being put on a constant diet. Given these results from two distinct cohorts, we assert that is reasonable to expect that a successful analysis of a set of ARISA profiles on the same samples should also generate a cluster that largely matches the cluster in Figure 2A where the time points cluster by subject.

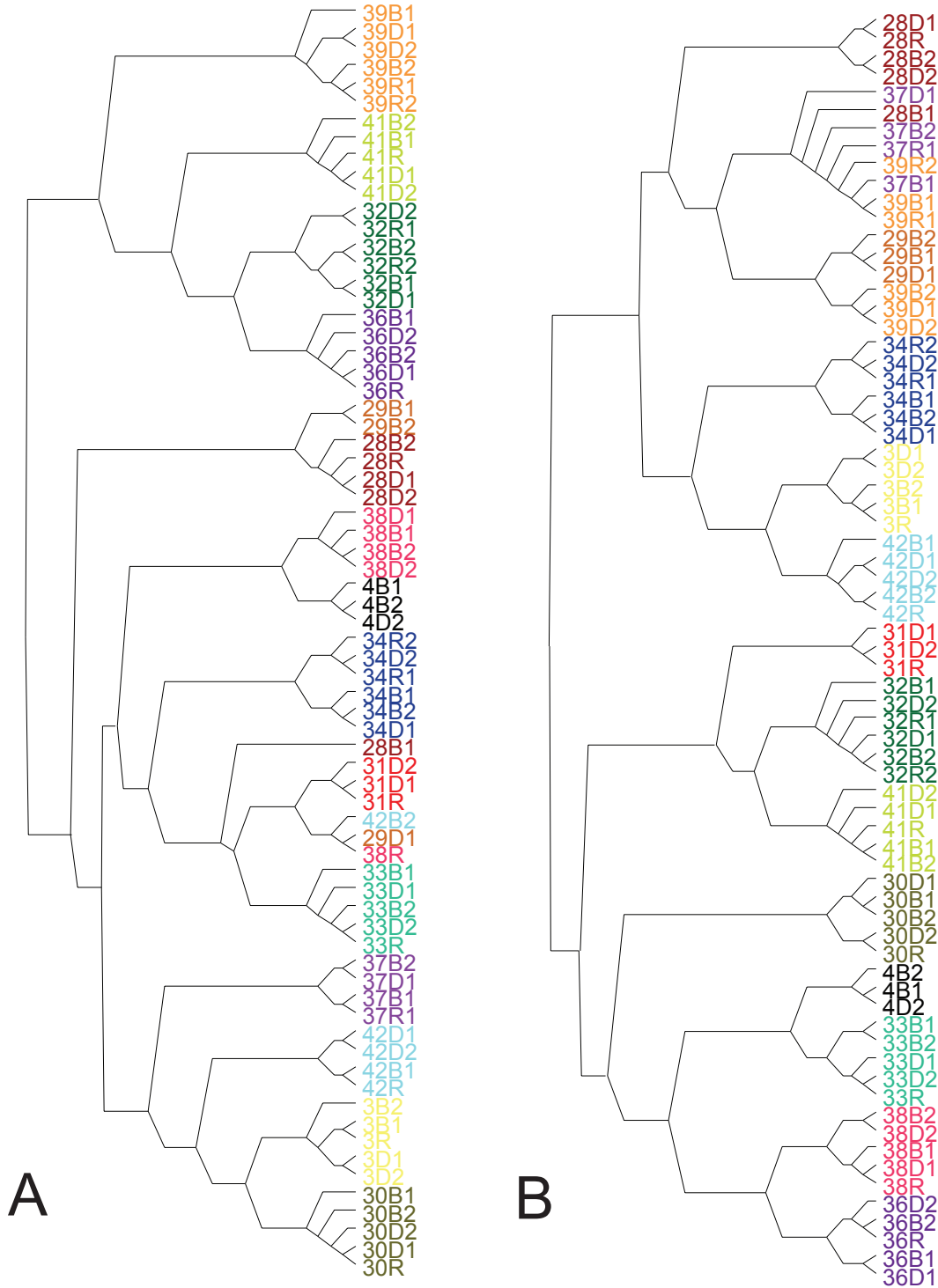

**Figure 2: Comparison of clusters using 454 next generation sequencing and ARISA.** (A) Hierarchical cluster of the 200 top OTUs of DNA sequences from the V1 region from 16S rRNA genes in microbial gut of human subjects via 454 sequencing. Clustering method = Wards. (B) Hierarchical cluster using Ward's clustering method on 71 ARISA profiles from human gut micro biome (value is fraction of total intensity, bin size = 3). Each color represents a different subject in the study (also noted by number). Time points

in the study are designated by the letter/number combo succeeding each subject number (B1 = baseline 1, B2 = baseline 2, D1=choline depletion 1, D2=choline depletion 2, R=choline repletion). So, for example, sample "39D1"represents subject 39 during the D1 (choline depletion) time point. See [15] for experimental details.

## Bin Size Strategies and Clustering Performance

In order to quantify the degree that samples from the sample subject cluster together over time point, we used the UNIFRAC distance metric [19,20]. In cases where samples could be perfectly discriminated by annotation, the UNIFRAC distance metric for a tree would be 1.0 indicating that no samples with distinct annotations shared close relatives. When analyzed with samples assigned by the subject id, the tree generated from our 454 sequences (Fig. 2A) has a UNIFRAC score of 0.995 indicating very strong, but not quite perfect, separation of samples by subject id. In our initial analysis of ARISA results, we asked whether the results of ARISA profiles could also generate such strong separation by subject.

For each of the samples for which we had 454 sequencing data, we also generated ARISA profiles (see methods). We clustered these 71 ARISA profiles using a variety of bin sizing strategies to determine how different analysis strategies would impact UNIFRAC score. The simplest of these binning methods is to group neighboring data signal into bins and assign an appropriate nucleotide length based on size standards (method 1 in Figure 3). Each bin represents different sized nucleotide fragments. Figure 2B depicts a tree generated using a bin size of 3 nucleotides using Wards clustering and normalizing the bins as fractions of total signal intensity. The UNIFRAC score for this ARISA derived tree was 0.95 demonstrating a very strong separation by subject that was nonetheless not quite as pronounced as the separation by subject observed for the sequence derived tree. To discover if the small inconsistencies between ARISA and 454-sequencing data could be explained by choices made during ARISA data analysis, we explored the effects of different combinations of analyses on clustering performance. We began with bin sizes. One possible limitation with the "simple bin" binning strategy that created Fig. 2B is that electropherograms are often observed to have minor shifts in the relative position of peaks when compared to one another. This can result in bin mismatches that should otherwise be the same, especially when bin sizes are smaller. Fisher and Triplett observed size variations of 1-2 nucleotides for fragments less than 1000 base pairs long and variations up to 13 nucleotides for larger DNA fragments [11]. To address these inconsistencies, larger bin sizes have been used to accommodate for separation variability and the loss of precision of larger fragments [11]. A bin size of 3 base pairs or larger can accommodate small shifts in the electropherograms. We refer to all methods that use a constant bin size across the electropherogram as "simple bins". A potential downside to these strategies is that as the bin size increases, there is a danger of grouping multiple peaks into a single bin (thereby losing resolution) and therefore we evaluated simple bin sizes ranging from 1 to 10 nucleotides in length. An additional danger to simple binning is that a peak signal could be split into 2 neighboring bins, falsely labeling neighboring bins at containing peaks.

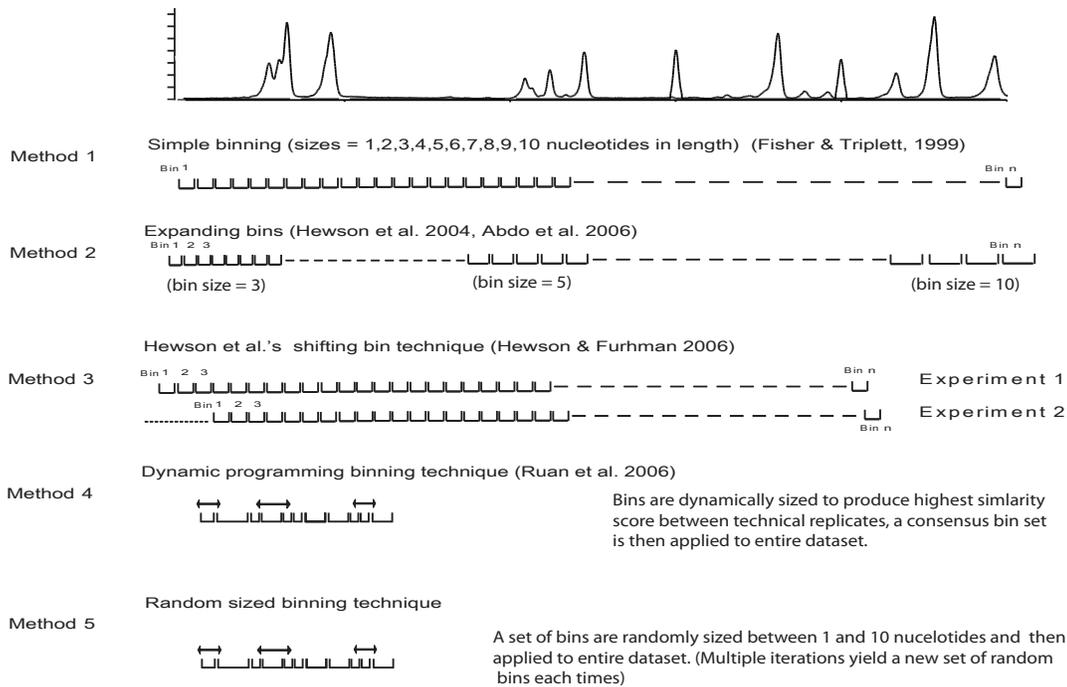

**Figure 3: Depiction of various binning methods used in ARISA cluster analysis.**

A variation on simple binning is to expand bin sizes for the larger DNA fragments to accommodate loss of reproducibility on separation (method 2, Figure 3). Since there is greater accuracy for smaller fragment lengths it has been suggested that bin size = 3 nucleotides for fragments less than 500, and bin size = 7 nucleotides for DNA lengths greater than 500 [21]. A binning strategy suggested by Abdo et al. utilized a bin size = 3 nucleotides from fragment sizes ranging from 400 to 700 base pairs, bin sizes = 5 nucleotides for fragments ranging 700-1000 base pairs and bin sizes = 10 nucleotides for fragments 1000-1200 base pairs [22]. In both methods larger bin sizes are used for longer DNA base pair lengths. These larger bins accommodate the more pronounced drift observed with longer DNA fragments, while still allowing high resolution for the smaller base pair lengths.

Since technical replicates are commonly run for quality control purposes, a further attempt to improve upon previous binning strategies was suggested by Hewson and Fuhrman [23] utilizing the technical replicates. They used a shifting bin strategy to minimize the differences observed in replicate profiles (method #3 in Figure 3) where an entire set of bins are shifted one nucleotide at a time and tested for similarity between replicates. Each replicate pair is compared by determining a distance metric where the differences within each bin are scored. Similar scoring bins will have smaller differences and therefore smaller overall distance scores. The bin shifting technique then shifts the data of one of the two replicates by a single nucleotide and then recalculates distance score for the replicate pair. This method repeats this shifting step for as many times as there are nucleotides in the largest bin, each time calculating scores until the best shift is found that minimize the distance score between the replicates. Once the best shift for each technical replicate pair is determined, the most commonly occurring best shift among all pairs is applied to the entire dataset prior to clustering. A potential weakness of this binning strategy occurs during the last step of the process, where the most common best performing shift is applied to all the datasets. The shift could adversely affect a small subset of the ARISA profiles that would have benefited from a different shift or no shift at all.

A more recently published ARISA clustering method implements a dynamic programming strategy for binning [24]. Instead of bins of a set size, Ruan et al. attempt to dynamically allocate the bin sizes across a set of profiles. This is done again by comparing replicate profiles to one another and selecting criteria that will yield the most similar results between the 2 replicates. For dynamic programming, bin sizes are varied on a per bin basis (ranging in bin sizes from 3 to 10 nucleotides) for each replicate pair. The best bin size is determined for every base pair position along the electropherogram (again determined by minimized scoring distance between replicate pairs). An ideal set of bin sizes is then selected by tracing back through the best bins. The dynamic programming portion of the algorithm involves determining bin scores that minimizes the Euclidian distance between 2 technical replicates and the subsequent trace back [24]. Method #4 in Figure 3 summarizes the dynamic programming binning method. Once the best bin sizes are determined for each replicate pair, a single composite profile of the most commonly occurring bin sizes in base pair space is then applied to all the profiles in the dataset.

To assess how well the different binning strategies perform, we developed a random binning strategy that creates a series of random bin sizes between 1 and 10 nucleotides in length (method #5 in Figure 3). This single set of randomly generated bins is then applied to the entire dataset. Unlike other binning methods discussed here, this method can be run multiple times generating a new set of bins each time that is then applied across all datasets. We ran each random binning method multiple times per condition and compared the results to the other binning methods with the null hypothesis that the random binning method performs as well as other binning strategies. We also applied two separate detection threshold criteria. The peaks in each ARISA profile were defined as either a fraction of total signal intensity (area under the curve) or as being present or absent from the bin (binary format). It has been demonstrated that the binary format can be more sensitive to peak detection [27].

Using the UniFrac score for each of the different bin assignment methods described above, we found that none of the methods were able to distinguish the different microbial communities as well as the cluster produced with our 454 dataset (Fig. 4). That is, while the UNIFRAC score were often high indicating strong separation by subject, no analysis path of ARISA data produced a UNIFRAC score as high as 0.995 (red dashed lines in Figure 4), the score we observed on our sequence derived tree. No matter which analysis path is taken, therefore, ARISA profiles therefore can approach, but never match, the strong separation by subject we observed with clustering derived from sequences

In order to determine if there were statistically significant differences between the performance of the binning methods, the random bin sizing method was performed 20 times and the average score and standard deviation was calculated. All binning methods were compared to the random binning scores. When data is normalized as fractions of total fluorescent signal, no binning method scored significantly better or worse than random binning after correcting for multiple comparisons ($P > 0.0019$, Bonferroni corrected at .05) (data not shown). When converting each of the bins into a simpler presence versus absence score (1 = peak, 0 = no peak), a slight increase in variability is seen amongst the various binning methods but again no method was significantly better or worse than random binning (Fig. 4C). We conclude that for our data set the choice of binning algrotihm has little overall effect on performance.

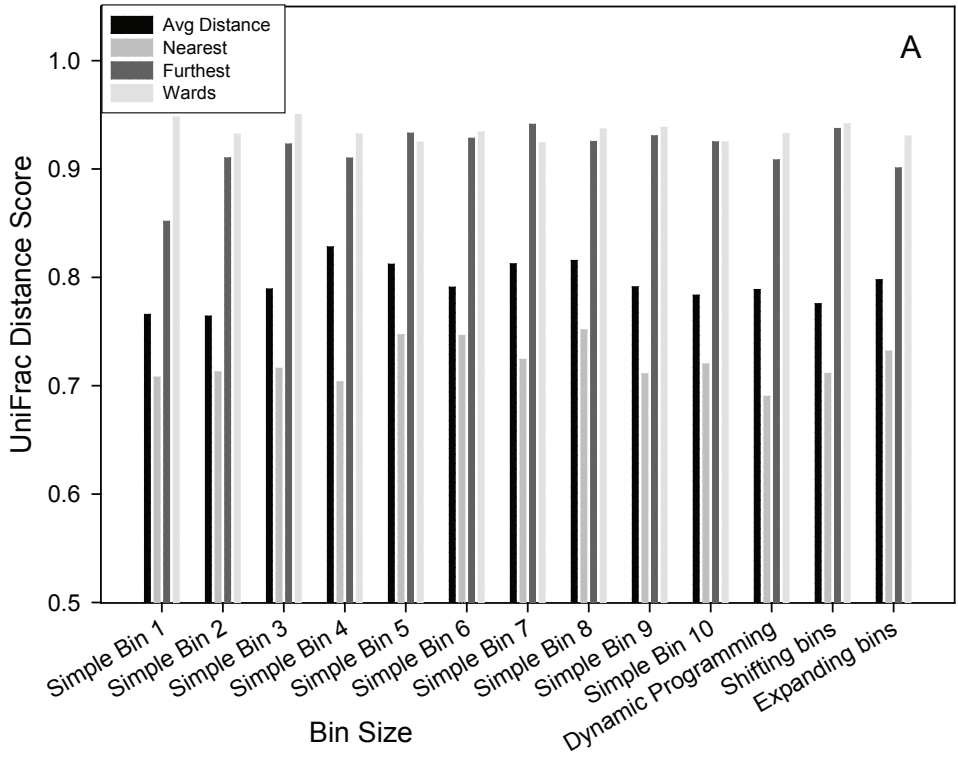

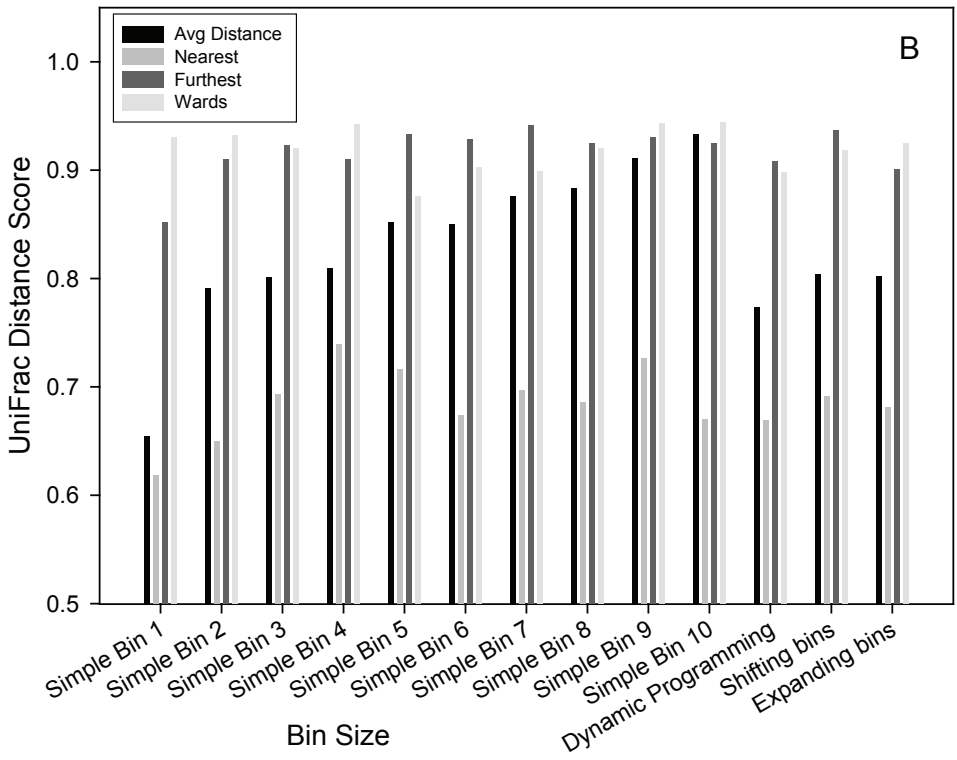

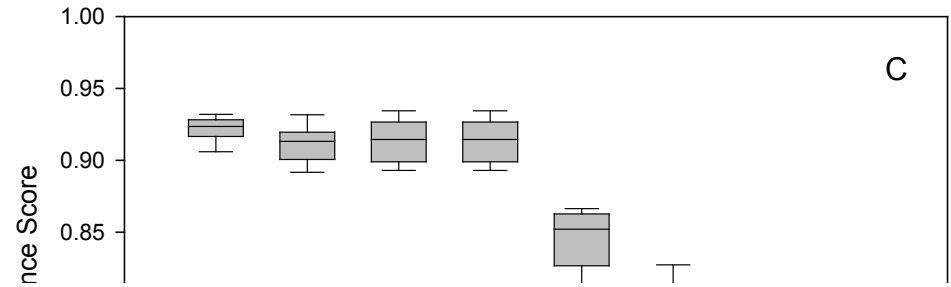

**Figure 4: Comparison of different clustering methods using UniFrac.** (A-B) Four clustering methods were compared using fraction of signal intensity (panel A) and presence/absence calls (Binary format, panel B) across different binning methods. Using the UniFrac metric, Wards cluster method performs best for the majority of binning methods, with furthest neighbor also performing well in most instances. Nearest Neighbor clustering method performs poorly regardless of bin size. Panel C: Comparison of random bins to other binning strategies (binary format). Panel D: For 20 iterations of random binning, average distance and nearest neighbor methods clearly yield poorer UniFrac distance scores with binary formatting contributing to a further decrease compared to the fraction of signal intensity format. Red dashed line in all panels represents the UniFrac distance score for the cluster of the 200 top OTUs of DNA sequences from the V1 region from 16S rRNA genes in microbial gut of human subjects via 454 sequencing.

In addition to UniFrac scores, the correspondence between sequencing and ARISA can be evaluated by simply counting the number of branches that failed to group with at least one other sample from the same subject. The smallest number of such events ranged from four (simple bin sizes 1 and 3) to at worst six (simple bin 5, 6 and 10) suggesting that while the differences between the various binning methods were minor, none of the binning methods matched the results seen by 454 sequencing where there were 2 such merged branches.

## Clustering methods

We next tested 4 clustering methods on the 71 ARISA profiles (Average distance, nearest neighbor, furthest neighbor in addition to Wards). Regardless of bin size chosen, the nearest neighbor algorithm generated UNIFRAC distances that were smaller than the other 3 clustering methods while the furthest neighbor and Wards methods produced consistently higher UNIFRAC distances. These differences persisted no matter whether the binning strategy was a published algorithm (Figs. 4A-4B) or random bins (Fig. 4D). The average distance method was worse across all binning methods when using a fraction of total signal intensity but did show better scores when using presence/absence format and larger bin sizes. However for random binning, the average distance method produced lower scores when using the presence/absence format.

## Discussion

The central assumption behind the scoring method used in our study is that different people have different gut microbial communities and these differences are stable over time despite a constant diet. This phenomenon has now been observed both in our previous study [15] and in a separate cohort analyzed with next-generation sequencing [18]. The stability of the gut microbial community over time is also consistent with a previous study in which samples taken several months apart from the same subjects were highly similar[16]. In order to evaluate ARISA data processing methods, therefore, we made the assumption that the strong separation shown in our sequence generated tree (Fig 2A), with a UNIFRAC score nearly 1, was "correct" and we evaluated whether any analysis path on our ARISA data could produce this high degree of clustering. It is possible, however, that the view of the samples provided by the ARISA is closer to biological "reality" and that our subjects had, in fact, more similar microbial communities. This possibility is a limitation of our study. However, we feel the more likely hypothesis is that the higher resolution of sequencing is yielding a more accurate view of the distinct microbial communities in different people than is achievable with the older PCR-based technology.

In this study we generated 360 tree dendrograms (4 clustering methods * 2 formatting methods * 13 bin strategies + 260 random binning trials) using 71 ARISA profiles and a variety of different parameters in an effort to create a tree that matches the result obtained by DNA sequencing. Using 454 DNA sequencing, the microbial communities cluster nearly perfectly by subject (Figure 2A). In the ARISA profiles, none of the 360 clustering results completely recapitulated the same separation of the subjects (Fig. 4). For the great majority of the 360 analysis paths we examined, the ARISA results did come reasonably close with as few as 5 "mis-clusters" in the tree (Figure 2B). Considering that ARISA is currently much less expensive than 454 sequencing, the fact that it generated nearly the same view as sequencing demonstrates that it remains a viable option for analysis.

Of all the decision parameters for this dataset, the choice of clustering method has the most impact with the Wards clustering method consistently outperforming other methods (Figure 4). Our finding of the superiority of the Ward's method is similar to the findings of Mangiameli et al. where Ward's cluster outperformed the majority of other hierarchical clustering methods tested [28]. Since Ward's clustering has a preference for smaller clusters when assigning nodes, however, the composition of our dataset is well suited to the Ward's method as there were at most 6 samples per subject. For other datasets that might contain larger clusters, Ward's tendency to favor small clusters might hurt its performance.

One obvious limitation in our analysis is that we used only a single dataset for testing from human stool samples. Microbial communities within stool samples are highly diverse and in this study we observed on the upper end, 60 varieties of species via ARISA (average number of peaks observed when bin size = 1, Table 2). If in our samples a bin choice occasionally inappropriately separated a peak into two separate bins in different samples, this had a small overall effect on clustering because our samples contained a sufficient number other peaks for comparison. Our results may not apply to less diverse ecosystems where issues such as bin splitting or shifting across samples may be more important and in such instances, the shifting bin and dynamic programming binning methods may therefore be more appropriate. On the other hand, for communities with a far greater diversity than human, such as an ocean metagenomic sample [29], we might expect that the high number of peaks would also saturate the bins minimizing the impact of the choice of binning algorithm. As always in analyses of complex datasets, it is crucial to understand the assumptions, strengths and limitations of each algorithm choice with regard to the dataset under consideration. Our results suggest that in cases where the microbial community is complex enough that bin saturation occurs, the binning choice is less likely to be important, freeing the user to maximize or minimize bins as appropriate for downstream analyses, or to repeatedly chose bins at random if a bootstrap estimate of error is required (as in Figure 4C in the current manuscript). On the other hand, if a sample contains only a few dominant peaks, it will be necessary to carefully choose a binning strategy that does not inappropriately split those few peaks across different samples.

**Table 2: Species richness table with increasing bin sizes. The average number of peaks detected across all ARISA profiles using different sized bins. Peaks for 10 random bins were also calculated (last row, average ± standard deviation)**

| Bin Size (nucleotide) | Number of peaks detected |
|---|---|
| Simple binning 1 | 60.1 |
| Simple binning 2 | 59.4 |
| Simple binning 3 | 56.2 |
| Simple binning 4 | 53.5 |
| Simple binning 5 | 50.1 |
| Simple binning 6 | 46.1 |
| Simple binning 7 | 43.6 |
| Simple binning 8 | 41.3 |
| Simple binning 9 | 38.6 |
| Simple binning 10 | 36.7 |
| Random binning | 43.3 ± 0.9 |

It has recently been pointed out that reproducibility is a central problem in fields such as environmental microbiology where "expensive or cutting edge techniques" are often used [30]. A barrier to performing sufficient replication is often cost. Our results demonstrate that some conclusions reached about community structure can be nearly replicated rapidly and inexpensively with older PCR-based technology and that the concordance between next generation sequencing and ARISA is insensitive to many choices made during ARISA analysis. If an observation is made by more than one technique, a high confidence can be assigned that the result is not an artifact of the chosen technique or an analysis pipeline associated with that technique. Techniques such as ARISA and T-RFLP, therefore, remain an important part of the scientific toolkit that can nicely complement high-precision next generation sequencing methods.

# Materials & Methods

## Sample Preparation

Microbial community analyses were performed as part of an NIH research (DK55965) study exploring the effects of common genetic polymorphisms that confer susceptibility to choline depletion [15]. Healthy female subjects (n = 15), a subset of those enrolled in a National Institutes of Health–funded study (DK055865) investigating choline metabolism, were recruited to participate in a gut metagenomic study and provided written and informed consent (approved by The Office of Human Research Ethics (OHRE) at the University of North Carolina at Chapel Hill). Stool samples were collected from fifteen human female subjects, who were hospitalized at the General Clinical Research Center (GCRC) of the UNC at Chapel Hill over a 60 day time course. The experimental design included placing subjects on diets that were strictly controlled and monitored for fat, carbohydrate and protein calories and for nutrients. Five to six fecal samples per subject were obtained at specific intervals during the study.

After human fecal samples were collected, they were shipped on dry ice to UNC Charlotte. DNA extraction from human fecal samples was performed using the Qiagen Stool Mini Prep kits. Approximately 180 to 220mg of human stool was measured for each subject per time point and bacterial DNA was extracted according to the Qiagen protocol. Approximately 180 to 220mg of fecal matter was measured for each subject per time point and bacterial DNA was extracted according to the manufacturer supplied protocol and then stored at -20 ºC until use. Additional details are described by Spencer et al. [15].

## ARISA Preparation

ARISA PCR was performed using universal bacterial primers 1406F-FAM (FAM+TGY ACA CAC CGC CCG T) and 125R (GGG TTB CCC CAT TCR G). Reactions were set up using 50ng of template DNA, estimated using a NanoDrop ND-1000 spectrophotometer (Thermo Fisher). Thermal cycling conditions were as follows: An initial denaturation step at 94°C for 2 minutes was followed by 35 cycles of 94°C for 25 seconds; 56.5°C for 30 seconds; 72°C for 60 seconds. Finally, an extension was carried out at 72°C for 5 minutes. Samples were loaded on an Applied Biosystems 3130 or 3130XL genetic analyzer. Applied Biosystems GeneScan™ 1200 LIZ® size standard was used to determine sizing up to 1200 nucleotides in length.

## 454 DNA Sequencing

The PCR products for 454 tagged sequencing were prepared with primers, reaction conditions, and thermal cycling parameters as described in Fierer et al. [31]. The 454 Life Sciences primer B with a "TC" linker and bacterial 27F primer (5'-GCCTTGCCAGCCCGCTCAGTCAGAGTTTGATCCTGGCTCAG-3') and 454 Life Sciences primer A with a "CA" linker, 12 mer barcode and bacterial primer 338R (5'-GCCTCCCTCGCGCCATCAGNNNNNNNNNNNNCATGCTGCCTCCCGTAGGAGT-3') were used to target the V1-V2 variable regions of the 16S rRNA gene. PCRs reactions used Platinum Taq DNA polymerase (Invitrogen) according to the supplier's protocol, with 100ng of bacterial genomic DNA as a template. Each reaction template was quantified using a PicoGreen assay (Invitrogen/Molecular Probes) on a NanoDrop ND-3300 fluorospectrometer (Thermo Fisher).  Samples were pooled in equimolar amounts and concentrated in a vacuum centrifuge before being submitted for 454 sequencing.

Operational taxonomic units (OTUs) were inferred from the sequences via the Ribosomal Database Project (RDP)-II infernal aligner and complete linkage clustering from the RDP web-based pipeline [32] (as described in [15]).

In addition to the 454 pyrosequencing, a small subset of samples was analyzed via Sanger sequencing targeting the 16S rRNA gene and the resulting DNA sequences were again clustered based on OTUs.  The Sanger sequencing OTUs confirm the 454 sequencing results in that the microbial communities identified cluster by subject and not by experimental condition over the 60 day time course (data not shown).

## Quality Control (QC) To Identify Poor ARISA Profiles

ARISA profiles were performed on aliquots of the same DNA used to generate samples submitted for 454 DNA sequencing. There were a total of 214 ARISA results including technical replicates. In analyzing these data, we used Peak Studio (http://fodorlab.uncc.edu/software), an open source alternative (paper currently under review) to the Applied Biosystems software for peak identification. A peak was defined as a positive slope over a set length of data points followed immediately by a negative slope of some length. A number of adjustable parameters were set that define the slope distances, inter peak distances between peaks, intra peak distances between slopes, peak lengths, and minimum peak heights. These peak calling parameters were manually adjusted so that the size standards within the ARISA profiles were optimally identified. Following peak identification of standards, all peaks in the ARISA spectra were identified with the same linear interpolation parameters and each ARISA peak was assigned a nucleotide length based off of neighboring size standards. The signal for each peak was ARISA peak area, normalized as a fraction of total signal.

 Each spectrum was manually inspected and those that still did not have the appropriate number of size standard peaks were excluded from further analysis. For the remaining datasets we applied a QC method that assessed how accurately size standards estimate nucleotide length by assigning nucleotide lengths using only every second size standard (i.e., only using half the size standards) and then predicting the length of each skipped size standard. The differences between the predicted skipped location (predictedSize) and the actual observed size (observedSize) were determined and the absolute sum of these differences was used to define a QC score (with a lower score being better).

$$\text{QC Score} = \sum_{i=1}^{N} | predictedSize_i - observedSize_i |$$

For technical replicates, we choose the profile with the best QC Score, which in the vast majority of cases was virtually identical to its replicate partner. We rejected any sample with a QC score above 1 (i.e. where the average error in predicting nucleotide size was > 1 nucleotide). At this threshold, we removed 61 of the 214 ARISA profiles to leaving 153 profiles available for clustering. Of these 153 profiles, 71 ARISA profiles were chosen in order to allow for a direct comparison between ARISA and the 71 samples used in 454 DNA sequencing.

**Technical Replicate Selection**

For purposes of quality control, ARISA experiments can be designed so that each ARISA profile is run in duplicate or triplicate (e.g., [14]) using the same DNA source as input into separate PCR reactions. By running replicates one can ensure technical consistency and if there are enough replicates, one can estimate variability when defining the intergenic fragment sizes. But it is not immediately clear how to use technical replicates in clustering analysis. Including all technical replicates can skew downstream analyses by violating the assumption of independence. For example, if a statistic is evaluating a null hypothesis that two environments have different ARISA profiles, that null hypothesis would likely be erroneously rejected if all technical replicates were included as independent samples. Treating technical replicates as an explicit factor in linear models would of course solve this problem, but when only two technical replicates are run per sample, there is an insufficient sample size to accurately estimate the within-group variance of technical replicates. For these reasons, therefore, it is often desirable to choose just one of the technical replicates to include in further analyses. We explored three different strategies for producing a single profile from multiple technical replicates. The first strategy involves selecting the best replicate based on the QC score. The second strategy averages two or more replicates together into one measurement prior to clustering, while the third strategy randomly selects one of the two technical replicates. We compared each of the three strategies by clustering the choline depletion study dataset using a bin size = 1, Ward's clustering method and each signal normalized as fraction of total signal intensity. We found that regardless of which technical replicate strategy was chosen, performance remained the same (data not shown). Choosing a technical replicate based on the best QC score or by averaging together two technical replicates offered no performance improvement over randomly picking a technical replicate for this dataset. The replicates with the best QC scores, as described above, were selected for subsequent study.

**Clustering Methods**

Four different clustering methods were applied to assess binning performance (average distance (UPGMA), nearest neighbors, furthest neighbors and the Wards clustering method [33]. The average distance method is the simplest way to generate distance measures between 2 branches. The distance (d), is determined by taking the absolute difference between each node in clusters x and y.

$$d(x,y) = \sqrt{\sum_i (x-y)^2}$$

The average of all the distances is then determined (average distance $= d(x,y)/N$).

In nearest neighbors clustering, the differences between cluster's x and y are again calculated, but the smallest distance between $x_i$ and $y_i$ is determined and used as the distance. Furthest neighbor clustering (also referred to as complete linkage clustering)[34] is identical to nearest neighbor, except that the largest distance between $x_i$ and $y_i$ is used for a distance.

Wards clustering method uses an analysis of variance type of approach to merge clusters. Cluster size is multiplied by an additional squaring step to calculate (d).

$$(d)_{wards} = \frac{n_x * n_y}{n_x + n_y} * \Delta centroid^2$$

$$\Delta centroid = \sqrt{\sum_i (x-y)^2}$$

The values $n_x$ and $n_y$ represent the number of branches contained in the cluster levels x and y. The purpose of Ward's is akin to ANOVA where the desired smallest distance is determined by first summing the distances (delta centroid), squaring the summed distance score and multiplying by $n_x$ and $n_y$. For two clusters equally far apart, the smaller cluster will be preferred for clustering. The clustering methods were implemented in java using a heavily modified version of ClusterLib, an open source implementation by Schulte et al. [35].

## Bin Strategy Comparison and Cluster Scoring Strategy

In order to determine the similarities between ARISA and DNA sequencing results, we utilized UniFrac to provide a quantifiable scoring metric to assess the influence of various ARISA parameters. UniFrac is a software tool that compares microbial communities based on phylogenic differences and determines if the communities are significantly different [19,25]. While UniFrac is usually performed on trees derived from 16S rRNA genes, the statistic can be applied to a phylogenetic tree derived from any source including binned ARISA results, where a distance matrix can be constructed using presence of peaks along the length of an electropherogram. UniFrac distance scores range from 0 to 1. A score of 1 represented a perfect separation between microbial communities, which in this study was defined as complete separation between subjects with no overlap. Scores approaching 0.5 showed little separation between the communities.

## Software Development

The code developed for analysis was written in Java 6.0. Each of the binning methods used was implemented in Java 6.0. Clusterlib was modified to analyze ARISA datasets (open source software available upon request). Tree viewing of clusters was performed using Archaeopteryx (http://phylosoft.org/archaeopteryx) [36], an open source phylogenetic tree viewer written in Java. UniFrac analyses were performed using a modified version of the UniFrac software [19,25] written in Python. Mantel statistics were performed using the "Analysis of Ecological Data" package (ade4) in R [37]. All java source code used is available upon request.

## Data availability

The original data file containing all pyrosequences has been submitted to the Short Read Archive at the National Center for Biotechnology Information under accession no. SRA012606.2 (http://www.ncbi.nlm.nih.gov/Traces/sra_sub/sub.cgi?&m=submissions&s=defaults).

## Acknowledgements

We like to give thanks to the lab of Stephen Zeisel for providing the resources for this project.